\documentclass[a4paper,fleqn,usenatbib]{mnras}

\usepackage[T1]{fontenc}
\usepackage{ae,aecompl}

\usepackage{graphicx}	
\usepackage{amsmath}	
\usepackage{amssymb}	

\title[Short title, max. 45 characters]{Discovery of pulsations from NGC\,300\,ULX1 and its fast period evolution}

\author[S. Carpano et al.]{
S. Carpano,$^{1}$\thanks{E-mail: scarpano@mpe.mpg.de}
F. Haberl,$^{1}$ 
C. Maitra,$^{1}$
G. Vasilopoulos$^{1}$
\\
$^{1}$Max-Planck-Institut f\"{u}r extraterrestrische Physik, Giessenbachstra{\ss}e 1, 85748 Garching, Germany}

\date{Accepted XXX. Received YYY; in original form ZZZ}

\pubyear{2018}

\newcommand{\xmm}{{\it XMM-Newton}}
\newcommand{\nus}{{\it NuSTAR}}
\newcommand{\swi}{{\it Swift}}
\newcommand{\cha}{{\it Chandra}}
\newcommand{\ngc}{NGC\,300\,ULX1}
\newcommand{\pdot}{$\dot{P}$}
\newcommand{\fdot}{$\dot{f}$}
\newcommand{\expo}[1]{$\times 10^{#1}$}

\newcommand{\msun}{$M_{\odot}$}
\newcommand{\ergcm}[1]{erg cm$^{-2}$ s$^{-1}$}
\newcommand{\ergs}[1]{erg s$^{-1}$}

\begin{document}
\label{firstpage}
\pagerange{\pageref{firstpage}--\pageref{lastpage}}
\maketitle

\begin{abstract}
The supernova impostor SN~2010da located in the nearby galaxy NGC 300, later identified as a likely supergiant
B[e] high-mass X-ray binary, was simultaneously observed by \nus\ and \xmm\ between 2016 December 16 and 20, 
over a total time span of $\sim$310\,ks.
We report the discovery of a strong periodic modulation in the X-ray flux with a pulse period of 31.6\,s and 
a very rapid spin-up, and confirm therefore that the compact object is a neutron star. 
We find that the spin period is changing from 31.71\,s  to 31.54\,s over that period, with a spin-up rate 
of $-5.56\times10^{-7}$\,s\,s$^{-1}$,  likely the largest ever observed from an accreting neutron star.
The spectrum is described by a power-law and a disk black-body model, leading to a 0.3--30\,keV unabsorbed luminosity of $4.7\times10^{39}$ erg\,s$^{-1}$. 
Applying our best-fit model successfully to the spectra of an  \xmm\ observation from 2010, suggests that the lower fluxes of \ngc\ reported from observations around that time are caused by a large amount of absorption, while the intrinsic luminosity was similar as seen in 2016. 
A more constant luminosity level is also consistent with the long-term pulse period evolution approaching an 
equilibrium value asymptotically.
We conclude that the source is another candidate for the new class of ultraluminous X-ray pulsars.

\end{abstract}

\begin{keywords}
stars: neutron -- pulsars: individual: \ngc\ -- galaxies: individual: NGC~300 -- X-rays: binaries
\end{keywords}



\section{Introduction}
\label{sec:intro}
NGC\,300 is a nearby, face-on spiral galaxy located in the Sculptor galaxy group at a distance of 1.88~Mpc \citep{2005ApJ...628..695G}. 
The X-ray source population of the galaxy was first studied using ROSAT data \citep{2001A&A...373..473R} and later with \xmm\ \citep{2005A&A...443..103C}. 
So far the brightest, but persistent, X-ray  source was the Wolf-Rayet/black hole X-ray binary NGC~300~X-1, with a luminosity of $\sim2\times10^{38}$ erg~s$^{-1}$ \citep{2007A&A...461L...9C}.
In May 2010 a new bright source appeared in NGC\,300, at optical wavelengths, that was originally classified as a supernova \citep{2010CBET.2289....1M}. The source was also detected by \swi\ XRT with an
unabsorbed luminosity of  $6\times10^{38}$ erg~s$^{-1}$ \citep{2010ATel.2639....1I}. Four months
later, SN\,2010da was detected at $\sim$7$\sigma$ significance by \cha\ ACIS-I  at coordinates $\alpha_\text{J2000}=00^\text{h}55^\text{m} 04\fs{}85$ and
$\delta_\text{J2000}=-37^\circ 41' 43\farcs 5$ with a 0.3--10\,keV unabsorbed luminosity of $\sim$2$\times10^{37}$ erg~s$^{-1}$ \citep{2011ApJ...739L..51B}. From the high X-ray luminosity associated to the large optical/IR flare, the authors conclude that the system is a high-mass X-ray binary (HMXB) system in outburst.
More recent \cha\ observations of the source were carried out in May and November 2014, when the luminosity changed by a factor of $\sim$10, going from $\sim$4$\times10^{36}$ erg~s$^{-1}$ to $\sim$4$\times10^{37}$ erg~s$^{-1}$ \citep{2016MNRAS.457.1636B}, which is much less than what was observed during the 2010 outburst. Using spectroscopic and photometric data performed in the UV to infrared wavelengths, the source has been later reinterpreted as a supergiant B[e] HMXB \citep{2016ApJ...830..142L,2016ApJ...830...11V}.

Ultraluminous X-ray sources \citep[ULXs; for a review see][]{2017ARA&A..55..303K} are off-nuclear point-like X-ray sources exceeding the (isotropic) 
Eddington limit for a neutron star or stellar-mass
black hole, which is $1.3\times10^{38} (M/$\msun$)$ erg~s$^{-1}$. For a 1.4\,\msun neutron star, the  Eddington limit (bolometric luminosity) is 
$\sim$2$\times10^{38}$ erg~s$^{-1}$. Several pulsating ultraluminous X-ray sources (ULPs) have been discovered in nearby galaxies, the brightest ones
being M82\,X-2, NGC\,7793\,P13 and NGC\,5907\,ULX1. M82\,X-2 was discovered with  \nus\  having an average pulse period of 1.37\,s
with a 2.5\,d sinusoidal modulation, and a 3--30\,keV luminosity of $4.9\times10^{39}$ erg~s$^{-1}$ \citep{2014Natur.514..202B}. The period derivative
$\dot{P}=-2\times10^{-10}$\,s\,s$^{-1}$ was calculated over a time interval of 5 days. The second bright ULP, NGC\,7793\,P13 has a pulse period of 0.42\,s
discovered with \xmm\ and \nus\ in May 2016 with a peak luminosity of $\sim$10$^{40}$ erg~s$^{-1}$ and a $\dot{P}=-3.5\times10^{-11}$\,s\,s$^{-1}$ measured over a time span of three years \citep{2016ApJ...831L..14F,2017MNRAS.466L..48I}. The third very bright ULP, NGC\,5907\,ULX1 with a bolometric luminosity reaching $10^{41}$ erg~s$^{-1}$, was discovered with a pulse period of 1.1\,s,  a $\dot{P}=-8\times10^{-10}$\,s\,s$^{-1}$ and a probable orbital modulation of 5.3\,d \citep{2017Sci...355..817I}.
Several other super-Eddington pulsating sources have been reported in the literature, one of these being CXOU J073709.1+653544, in NGC\,2403, which is one of the fastest spin-up ULPs with a $\dot{P}=-1.1\times10^{-7}$\,s\,s$^{-1}$ with a pulse period of $\sim$18\,s \citep{2007ApJ...663..487T}.

In this letter, we report the discovery of the pulse period of the supernova impostor SN\,2010da, hereafter referred as \ngc, observed in a very long \xmm\ observation performed simultaneously to \nus\ in December 2016. The discovery was already announced via the ATel$\#$11158 \citep{2018ATel11158....1C}.  The first section describes the \xmm\ and \nus\ observation and data reduction. We then explain in Sec.~\ref{sec:time} our analysis of the pulse period and the period derivative. A spectral analysis of the source is performed in Sec.~\ref{sec:spec}. Conclusions are ending the paper.

\section{Observations and data reduction}

\ngc\ was visible in three \xmm\ \citep{2001A&A...365L...1J} observations: 0656780401 performed on 2010 May 28 for a duration of 18\,ks when the source was observed in outburst for the first time, and two consecutive observations, 0791010101 and 0791010301, performed from 2016 December 17 to 20, for a duration of 139+82\,ks. The data were reduced following standard procedures using the \xmm\ SAS data analysis software version 16.1.0 with the calibration files (CCFs) available in January 2018. 
For the source extraction region we used a circle around the \cha\ coordinates mentioned in Sec.~\ref{sec:intro} with a radius of 40$''$. The background region was
centred on a region equidistant from the source extraction region to the nearby source NGC\,300\,X-1, in order to minimise possible contamination after background subtraction. It was centred on $\alpha_\text{J2000}=00^\text{h}55^\text{m} 07\fs{}50$ and $\delta_\text{J2000}=-37^\circ 43' 15\farcs 3$ with a radius of  40$''$.
 For the spectral analysis, periods of flaring background were removed from the data by removing the time intervals with background rates $\geq$ 8 and 2.5 cts ks$^{-1}$ arcmin$^{-2}$ for EPIC-pn and EPIC-MOS 
respectively \citep{2013A&A...558A...3S}. Event extraction was performed using the SAS task \texttt{evselect} applying the standard filtering criteria (\texttt{\#XMMEA\_EP \&\& PATTERN<=4} for EPIC-pn and \texttt{\#XMMEA\_EM \&\& PATTERN<=12} for EPIC-MOS).

The source was observed for the first time with \nus\ \citep{2010SPIE.7732E..0SH} on 2016 December 16, for an exposure time of 163\,ks. The data were processed using the  \texttt{NuSTARDAS} software version 1.8.0 (released with \texttt{HEASOFT} v.6.22.1) and calibration files (CALDB) version 20171002, while the spectra, response files and barycenter-corrected event files were produced using the \texttt{nuproducts} task. The source and background extraction regions are a circle and annulus, respectively, both centred on the \cha\ coordinates (since NGC\,300\,X-1 was not visible in the data), with a radius of 50$''$ for the source and radii of 60$''$ and 100$''$ for the background.

\section{Pulse period and its evolution}
\label{sec:time}

To refine the spin period evolution reported from \ngc\ \citep{2018ATel11158....1C}, 
we used a Bayesian method which uses an epoch-folding algorithm to calculate frequency-dependent odds-ratios,
which describe how the data favour a periodic model over the unpulsed model for a given frequency
\citep{1996ApJ...473.1059G,2008A&A...489..327H}. 
The analysis of the peak in the odds-ratio periodogram provides the most probable frequency and the 1$\sigma$ uncertainty. 

\begin{figure}
  \resizebox{0.95\hsize}{!}{\includegraphics[angle=-90,clip=]{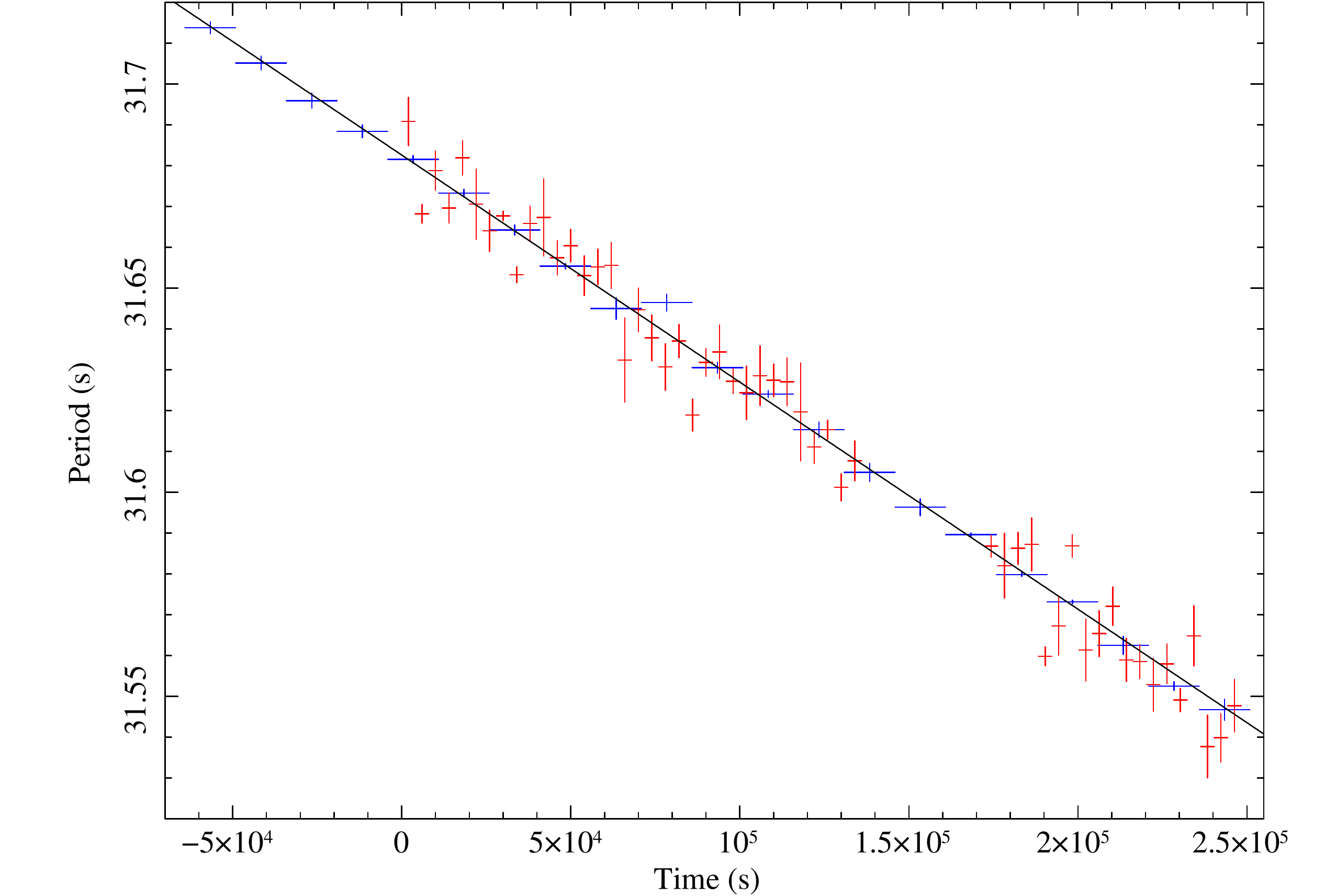}}
  \caption{
    Spin period evolution of \ngc\ obtained from 4\,ks intervals of EPIC-pn (red crosses) and 15\,ks intervals 
    of \nus\ data (blue crosses). The straight line represents the best-fit model of a linear period decrease applied 
    to both data sets. Time zero corresponds to the start of the EPIC-pn exposure.
  }
  \label{fig:pevolution}
\end{figure}

To investigate the period evolution during the \xmm/\nus\ observations we split the EPIC-pn data, which provides the best 
statistics with high time resolution (73\,ms), into 4\,ks intervals. The arrival times of the events from the source region 
(in the 0.2$-$10\,keV band) were then analysed for all 53 intervals from the two observations. 
For \nus\ we combined the data from the two instruments, used 21 intervals of 15\,ks and an energy band of 3$-$20 keV.
The inferred evolution of the spin period is shown in Fig.~\ref{fig:pevolution}. The spin period of \ngc\ decreased linearly from 
$\sim$31.71\,s at the start of the \nus\ observation to $\sim$31.54\,s at the end of the \xmm/\nus\ observations.
The period derivative inferred from a model with a constant and linear term fitted to the \xmm\ and \nus\ data is 
($-$5.563$\pm$0.024)\expo{-7}\,s s$^{-1}$ with a spin period of 31.68262$\pm$0.00036\,s 
at the start of the of the EPIC-pn exposure (MJD 57739.39755).  
We note that the introduction of an additional sine function to fit the period evolution seen from EPIC-pn as reported in \citet{2018ATel11158....1C} 
was caused by a software problem. In our refined analysis, we didn't find any significant deviation from a linear model and derive 
an upper limit for v$\cdot$sin(i) of $\sim$4.6\expo{-5}c (assuming orbital periods between one and three days), 
which translates into upper limits for the mass function of $\sim$8\expo{-4}\msun\ and for the inclination of $\sim$3\degr\ for a 20\msun\ companion star.

\begin{figure}

\resizebox{\hsize}{!}{\includegraphics[clip=]{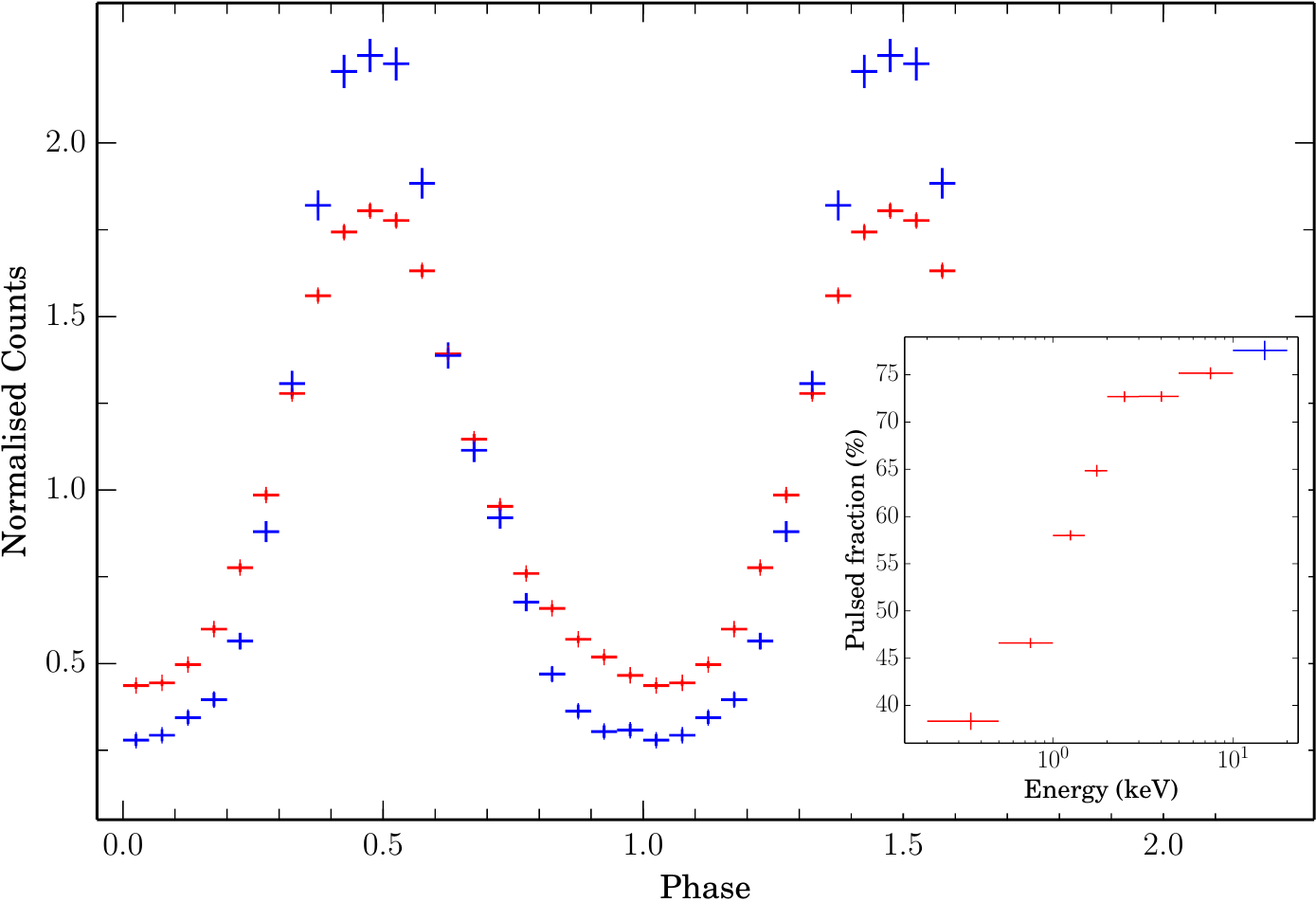}}
  \caption{
    Pulse profile of \ngc\ obtained from the 0.2$-$10\,keV EPIC-pn data of the first 2016 observation (red),
    and the 3$-$20\,keV light curve from the combined \nus\ instruments (blue). Both profiles are background subtracted
    and corrected for the pulsar spin-up. The inset shows the pulsed fraction as function of energy with colours 
    as in the main plot.
  }
  \label{fig:pulse}
\end{figure}

Taking into account the constant value for \pdot\ derived above, we converted the arrival times to spin phase
($\phi = \phi_0 + (t-t_0) \times f + 0.5(t-t_0)^2 \times $\fdot).
The pulse profiles (phase histogram) for the first \xmm\ and the \nus\ observation are shown in
Fig.~\ref{fig:pulse}. The pulsed fraction (0.2$-$10\,keV), defined as proportion of 
flux integrated over the pulse profile above minimum flux relative to the total integrated flux, 
increased slightly from 56.3$\pm$0.3\% during the first 2016 \xmm\ observation to 57.4$\pm$0.3\% in the second. 
The pulsed fraction also increases strongly with energy with 72.1$\pm$0.4\% in the \nus\ data (3$-$20\,keV).

\section{Spectral analysis}
\label{sec:spec}
X-ray spectral analysis was performed using {\small XSPEC} version 12.9.1 \citep{1996ASPC..101...17A}.
To account for the interstellar absorption, the model \emph{TBabs} was used, with abundances from 
\cite{2000ApJ...542..914W}. In order to investigate the broad-band X-ray spectrum of \ngc, 
we fitted the spectra from the overlapping 2016 \xmm\ and \nus\ observations simultaneously. 
Spectra from all the EPIC cameras and \nus\ focal plane modules (FPMA and FPMB) were fitted together with a 
constant multiplicative factor accounting for inter-calibration uncertainties and an intensity change between the \xmm\ observations
(with respect to EPIC-pn of 0791010101). The \nus\ spectral analysis was limited 
at high energies to 30 keV due to the background dominating the spectrum above this energy.

\begin{figure}
  \begin{center}
  \resizebox{0.94\hsize}{!}{\includegraphics[clip=,angle=-90]{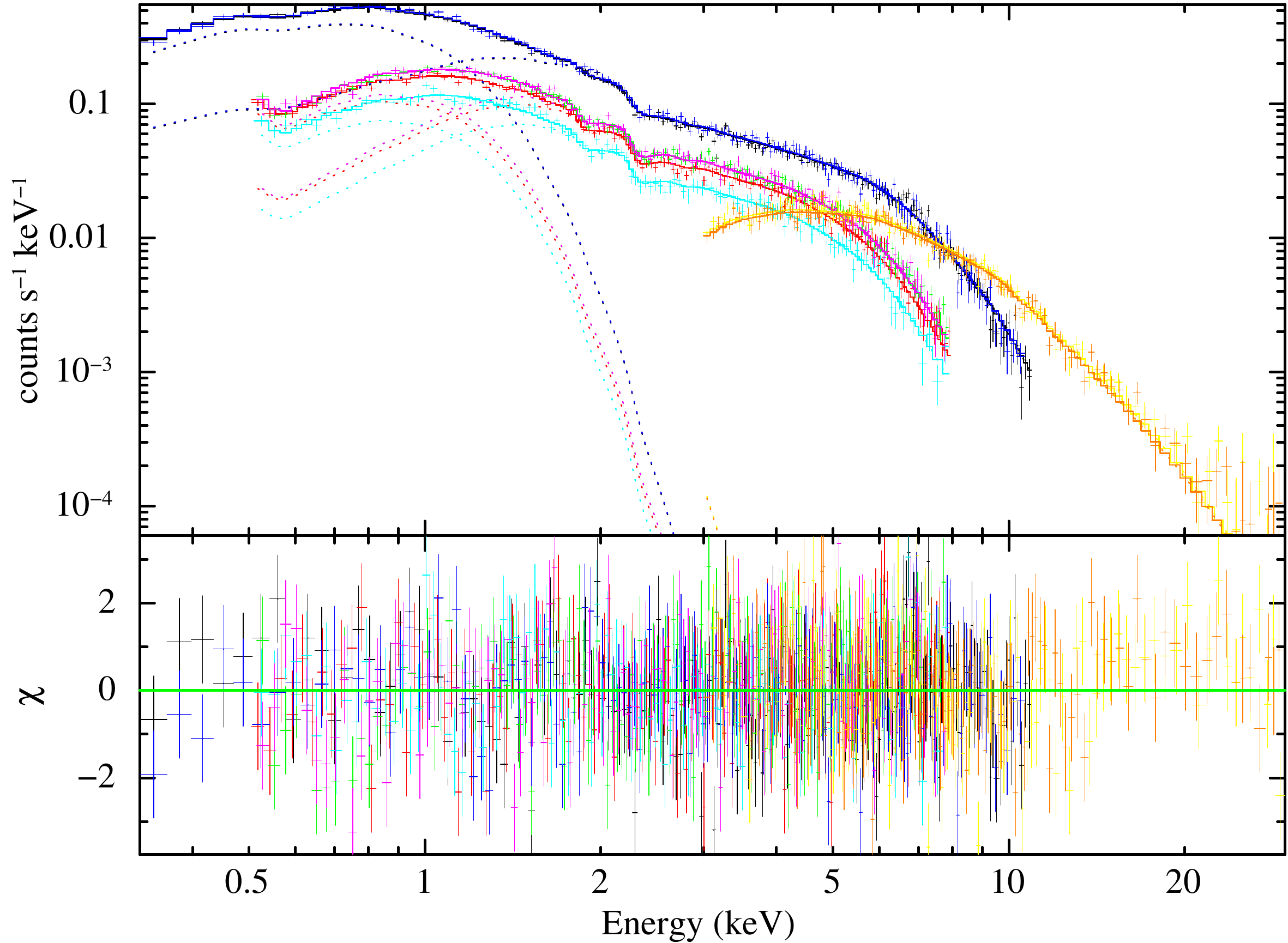}}
  \resizebox{0.94\hsize}{!}{\includegraphics[clip=,angle=-90]{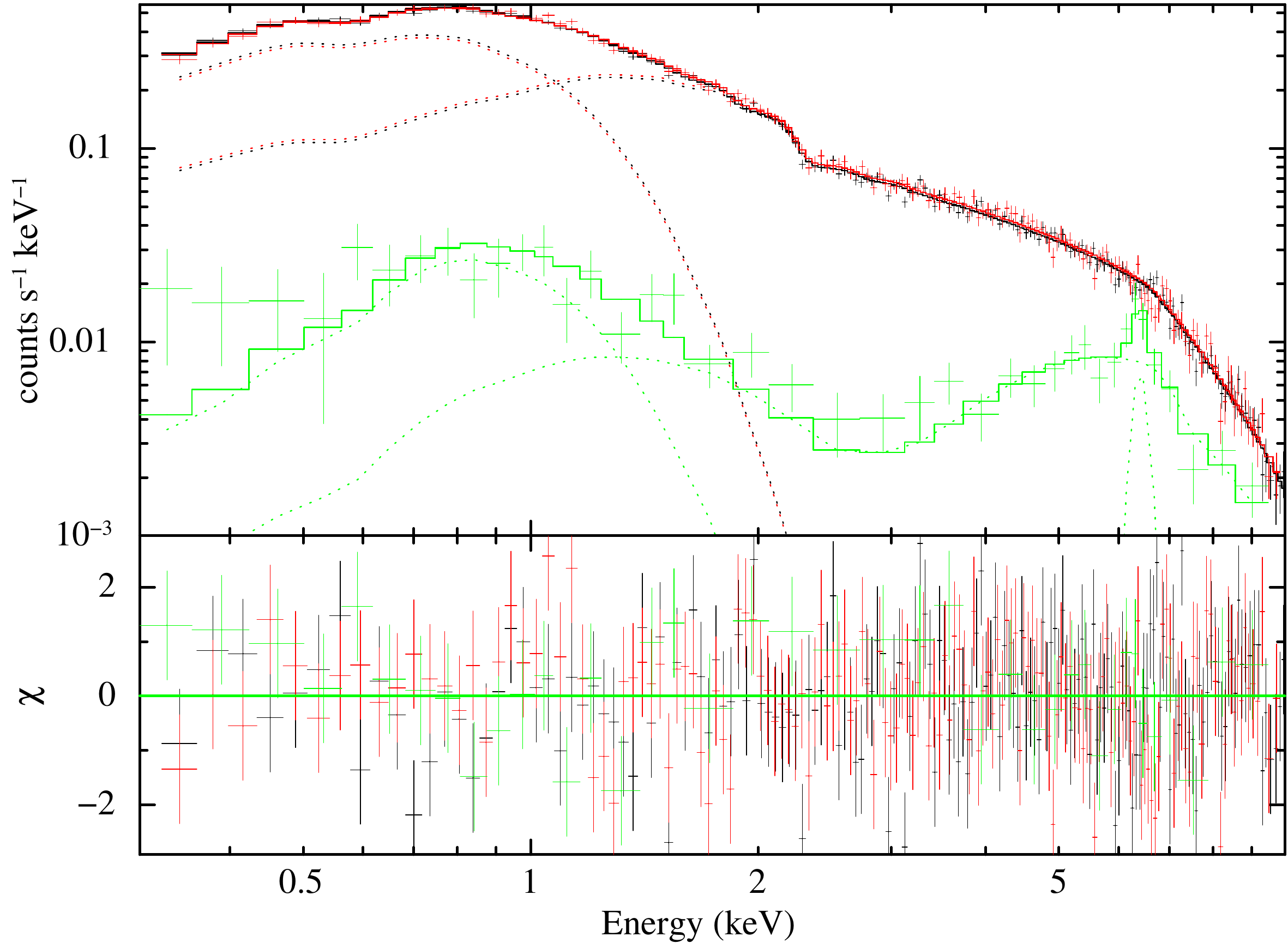}}
  \end{center}
  \caption{The top panels show the simultaneous broad-band spectral fit of \ngc\ using the 2016 \xmm\ and \nus\ spectra from all the 
           \xmm/EPIC cameras (pn in black and blue, MOS in red, green, magenta and cyan) and \nus\ focal plane modules 
           (FPMA and FPMB in yellow and orange), along with the best-fit model (upper panel). The panel underneath displays the 
           residuals for the best-fit model. 
           The bottom panels show the simultaneous spectral fit of \ngc\ using the EPIC-pn spectra together with the residuals as above. 
           Observation 0791010101 is marked in black, 0791010301 in red and 0656780401 (from 2010) in green.}
   \label{spec}
\end{figure}

We started fitting the spectra with a two-component model consisting of a power law with high-energy cutoff and a soft thermal 
component (disk black-body).
Similar models have been successfully used to model the spectra of ULXs hosting neutron stars \citep{2017MNRAS.466L..48I,2017ApJ...836..113P}. 
However, the resultant fit to the spectra of \ngc\ was unsatisfactory, with residuals visible at energies $<$0.5\,keV. 
The residuals indicate the presence of a further softer spectral component which can be attributed to the scattering 
and reprocessing of the X-ray photons originating in the vicinity of the neutron star, by an additional absorber. 
This was modelled using a partial-covering absorber component. 
Further addition of this partial absorber component to the power law alone did not improve the fit. However, the fit improved 
significantly ($\Delta\chi^{2}=20$) and was satisfactory when the partial absorber was applied to the power-law and black-body 
components together. This can be understood if the underlying continuum consists of a combination of power-law component 
(originating from the vicinity of the neutron star) plus a  disk black-body component (originating from the inner accretion disk), 
modified by scattering and absorption by additional material. This is most likely located in the clumpy wind of the supergiant companion or 
inner part of the circum-stellar disk of a Be star.
The reduced $\chi^{2}$ of the fit is 1.13 for 1167 degrees of freedom. The spectra and the best-fit model are shown in Fig.~\ref{spec} (top) 
and the spectral model with the best-fit parameters is summarised in Table. \ref{table1}.

\begin{table}
\caption{Simultaneous fit of the 2016 \xmm\ and \nus\ spectra.
Errors are quoted at the 90\% confidence.}
\begin{tabular}{c c c}
\hline \hline
Component & Parameter & Value \\
 \hline
TBabs & $\ensuremath{N_{\mathrm{H}}}$ $^a$ & 0.11 $\pm$ 0.01 \\
Pcfabs & $\ensuremath{N_{\mathrm{H}}}$ $^a$ & 0.75 $\pm$ 0.07 \\
       & cv$_{frac}$ & 0.85 $\pm$ 0.03 \\
Power law & $\Gamma$ & 1.52 $\pm$ 0.03 \\
         & Ecut (keV) & 5.6 $\pm$ 0.2 \\
         & Efold (keV) & 7.0 $\pm$ 0.3 \\
         & F$_{\rm x}$ $^b$ & $ 6.4 \pm 0.1 $\\
Disk black-body & \emph{kT} (keV) & $ 0.178^{+0.008}_{-0.007} $\\
 & F$_{\rm x}$ $^b$ & $ 4.4 \pm 0.1 $\\
 \hline
L$_{\rm x}$ $^c$ & & $ 4.7$\\
$\chi^2_\textrm{red}$/d.o.f. & & 1.13/1167\\
 \hline
\end{tabular}\\
\flushleft{
Spectral model used in \textsf{XSPEC} convention: \emph{TBabs}$<$1$>$*\emph{pcfabs}$<$2$>$(\emph{powerlaw}$<$3$>$*\emph{highecut}$<$4$>$ + \emph{diskbb}$<$5$>$+\emph{gaussian}$<$6$>$)\\
The obtained values of the normalisation constants for EPIC-MOS1 and MOS2 were $0.98 \pm0.01$ and $0.98 \pm0.01$ for 0791010101, and $1.01 \pm0.02$ and $0.99 \pm0.01$ for 
0791010301, respectively. The corresponding values for EPIC-pn for 0791010301 and \nus\ FPMA and FPMB were $1.05 \pm0.01$, $0.86 \pm0.02$ and $0.86 \pm0.02$, respectively.\\
$^{a}$ in $10^{22}$ atoms cm$^{-2}$\\

$^{b}$ Unabsorbed flux (0.3--30 keV) in  $10^{-12}$ \ergcm\\

$^{c}$ Unabsorbed luminosity (0.3--30 keV) in  $10^{39}$ \ergs\\
}
\label{table1}
\end{table}

The \xmm\ spectrum taken in 2010 is drastically different compared to that in 2016, with a soft component seen at energies $<$2 keV,
an almost flat spectrum between 2--4 keV and a bump-like feature above 5\,keV. This can be explained if the column density 
was significantly higher in 2010 and the direct component of the emission was reduced drastically. A similar approach was adopted
by \cite{2015MNRAS.447.2387C} to explain the change in the spectrum of the Be/X-ray binary pulsar SXP\,5.05.
To investigate this, we fitted the EPIC-pn spectra of the three observations simultaneously assuming the same
underlying continuum spectrum as used in the broad-band spectral fit, and allowing only certain parameters to vary. Apart from varying all the absorption components, 
the power-law normalisation was left to vary in order to account
for an intrinsic variation in the X-ray luminosity of the source. The resultant fit is satisfactory with a reduced $\chi^{2}$ of 1.06 for 345 degrees of freedom. The spectra and the best-fit model are shown in Fig.~\ref{spec} (bottom) 
and the spectral model with the best-fit parameters is summarised in Table. \ref{table2}. 
The change in the 2010 \xmm\ spectra can be explained by a large increase in the column density of both absorption components and the direct component of the underlying continuum reduced to a mere 5\%. 
This model predicts an absorption corrected intrinsic luminosity of 2.8$\times 10^{39}$ \ergs,
which is not very different from the luminosity estimates from the 2016 observations (Table \ref{table2}). 
An Fe-K$\alpha$ fluorescence line was noticed in the spectrum of the 2010 \xmm\ observation as expected in the case of reprocessed
emission. The line energy is $6.40_{-0.07}^{+0.14}$~keV and the equivalent width is  $348 \pm93$ eV. No line was required to fit 
the spectra of the 2016 observations as the continuum completely dominates at this energy range.

\begin{table}
  \caption[]{Results of the simultaneous fit to the EPIC-pn spectra from the 2016 and 2010 observations.}

    \begin{tabular}{lccc}
    \hline\hline
    Observation & 0791010101 & 0791010301 & 0656780401 \\
    \hline
    $\ensuremath{N_{\mathrm{H,TBabs}}} (10^{22})$  & 0.11 $\pm$ 0.02& 0.11$^{b}$&0.61$_{-0.12}^{+0.17}$\\
    $\ensuremath{N_{\mathrm{H,Pcfabs}}}  (10^{22})$  & 0.66$\pm$0.10& 0.66$^{b}$& 47.2$\pm$15.3\\
    cv$_{frac}$ & 0.84$\pm$0.04& 0.84$^{b}$&  $0.95 \pm 0.02$\\
    $\Gamma$ &  1.54$\pm$0.04 &  1.54$^{a}$&  1.54$^{a}$\\
    Ecut (keV)&  6.6 $\pm$ 0.3 &  6.6$^{a}$&  6.6$^{a}$\\
    Efold  (keV)&  4.8 $_{-0.8}^{+1.2}$ &  4.8$^{a}$& 4.8$^{a}$\\
    F$_{\rm x,pl}^{c}$ & $5.62 \pm 0.01$& $6.02 \pm 0.01$& $2.50 \pm 0.14$\\
    \emph{kT} (keV) & $0.17 \pm 0.01 $&  $0.17^{a}$  & $0.17^{a}$\\
    F$_{\rm x,bbody}^{c}$ & $5.60\pm 0.16$& 5.60$^{a}$ & 5.60$^{a}$\\
    \hline
    L$_{\rm x,total}^{c}$ & 4.2 & 4.4&  2.8 \\
    $\chi^2_\textrm{red}$/d.o.f. & & 1.06/345\\
    
   \hline
    \end{tabular}

  \flushleft{
  $^{\rm (a)}$ parameters of the underlying continuum were linked to the parameters obtained from observation 0791010101.\\
  
  $^{\rm (b)}$ The absorber components of the 2016 \xmm\ observations were linked as no significant variation was seen.\\
  
  $^{\rm (c)}$ Units for the unabsorbed flux and luminosity are as in Table~\ref{table1}, but in the energy range of 0.3--10 keV.
  }
 \label{table2}
\end{table}

\section{Discussion and conclusions}

\begin{figure}
 
\resizebox{\hsize}{!}{\includegraphics[clip=,angle=-90]{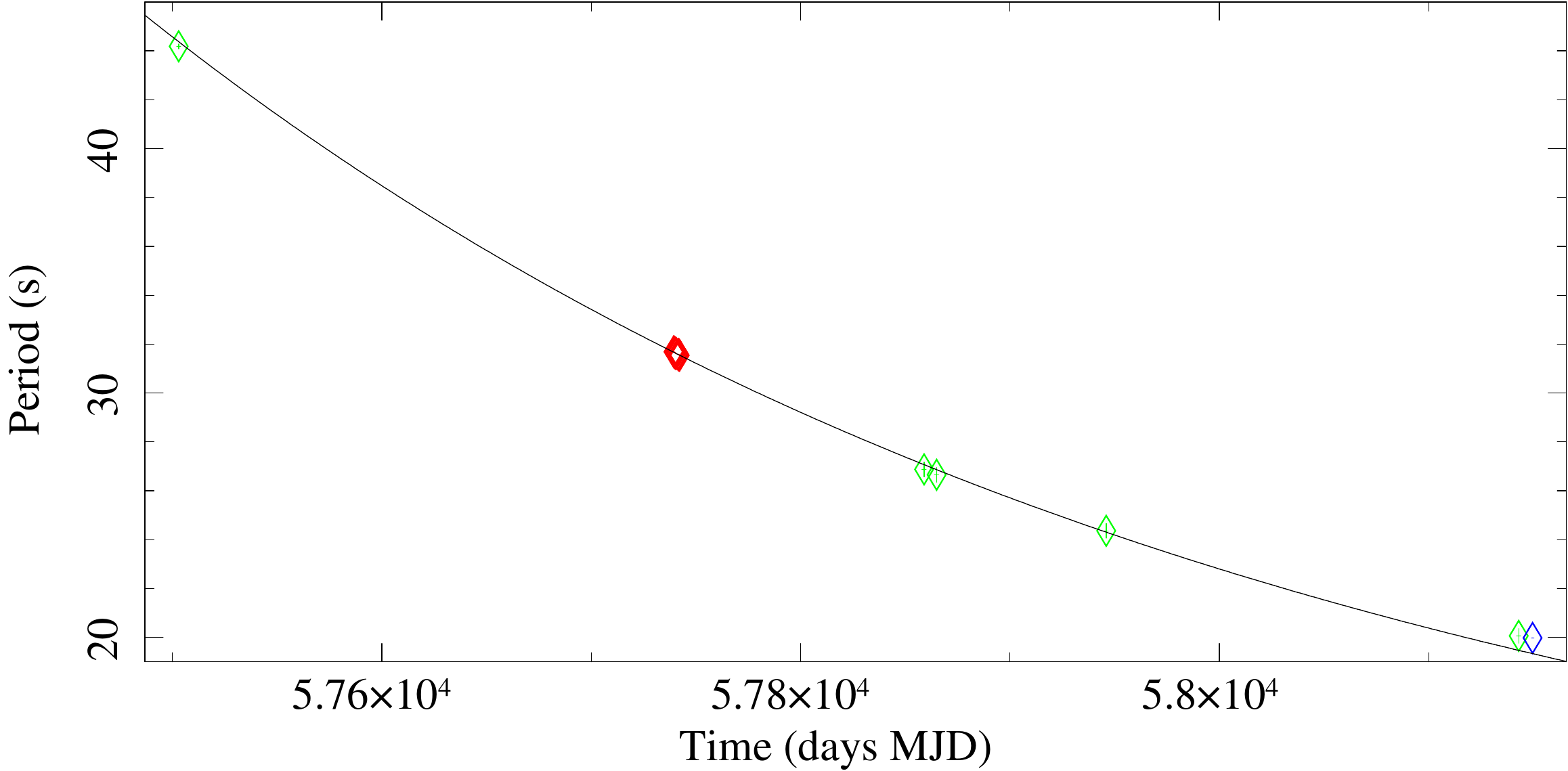}}
  \caption{
    Long-term time evolution of the pulse period of \ngc\ from 2016-04-25 to 2018-02-01.
    The simultaneous \xmm/\nus\ observation is marked in red, \swi\ in green and the recent \nus\ observation
    \citep{2018ATel11282....1B} in blue. To account for multiple solutions in the short \swi\ observations we
    use a $\pm$0.3\,s error which includes aliasing effects due to the \swi\ satellite orbit.
  }
  \label{fig:pltevo}
\end{figure}

We report the discovery of pulsations from \ngc\ with a period of $\sim$31.6\,s and strong spin-up during 
simultaneous \xmm/\nus\ observations.
To our knowledge, the secular spin period derivative of $-$5.56\expo{-7}\,s s$^{-1}$ seen over three days 
is the highest ever observed from an accreting neutron star. 
The analysis of archival \swi\ observations has shown that the strong spin-up seen during the \xmm/\nus\ 
observations lasts already for at least 21 months \citep{2018ATel11179....1V,2018ATel11228....1G,2018ATel11229....1K}.
The period history as shown in Fig.~\ref{fig:pltevo} follows an exponential decrease with e-folding time of $\sim$1.5\,years, 
approaching an equilibrium value asymptotically. 
A fit using a model with constant plus exponential function yields values between 8.5\,s and 14\,s, 
depending on the consideration of all measurements or only \xmm/\nus\ and later, respectively. 

The extreme spin-up of the pulsar with luminosity seen over the duration of the \xmm/\nus\ observations can be used to constrain the magnetic field
strength (B) of the neutron star. Using the standard disk accretion model of \cite{1979ApJ...234..296G}, B can be estimated
from the derived spin period, \pdot\ and the luminosity estimated from the \xmm/\nus\ observations. The dimensionless accretion torque
$\eta(\omega)$ is estimated using the analytic expression from \cite{1979ApJ...234..296G} which is accurate up to 5\% and is valid in our regime ($0 <\eta(\omega)<0.9$).
The estimated B is $\sim$3$\times10^{12}$ G. On the other hand the equilibrium spin period of $\sim$8.5$-$14\,s derived from Fig.~\ref{fig:pltevo}
can also be used to estimate B, for a given average luminosity \citep[see][]{2014MNRAS.437.3664H}. Assuming an average luminosity of $2\times10^{39}$ \ergs,
B is $\sim$1$\times10^{13}$ and $\sim$2$\times10^{13}$ G, assuming equilibrium spin period of 8.5 and 14\,s respectively. Although the estimate of B from the standard
accretion torque theory is lower than that predicted from the equilibrium spin period, the difference can be bridged by a considerable extent by accounting for a higher
accretion efficiency than obtained by considering the gravitational potential for a neutron star of standard mass (1.4 $M_{\odot}$) and radius (10 km). 
In any case these estimates need to be handled with caution as the model assumptions (geometrically thin disk and sub-Eddington accretion) are not fulfilled in 
the extreme case of \ngc.

The average pulsed fraction seen from \ngc\ of around 55\% in the \xmm\ energy band 0.2$-$10\,keV and 72\% in the \nus\ band 3$-$20\,keV is higher 
than for the other known ULPs: 
for NGC\,7793\,P13 it increases from $\sim$10\% to $\sim$40\% in the \xmm\ band \citep{2017MNRAS.466L..48I},
for NGC\,5907\,ULX1 it grows from $\sim$12\% to $\sim$20\% \citep[\xmm\ band]{2017Sci...355..817I},
and M82\,X-2 also shows the highest pulsed fraction of $\sim$23\% at higher energies \citep[10$-$30 keV]{2014Natur.514..202B}. 
The pulsed fraction for \ngc\ rises strongly from $\sim$37\% (0.2$-$0.5\,keV) to $\sim$70\% around 2\,keV. 
Above 2 keV the rise is flatter reaching nearly 80\% in the 10$-$20\,keV band. This suggests that the soft black-body like emission component is 
not pulsed and does not originate at the neutron star surface. A likely origin is the accretion disk. 

Our analysis of the X-ray spectra obtained by \xmm\ and \nus\ in Dec. 2016 shows that they can be represented 
by a two-component model with a dominating power law (photon index $\sim$1.6 and soft black-body emission (kT $\sim$0.18\,keV).
A high-energy cutoff around 6.6\,keV is similar as seen from other ULXs \citep{2009MNRAS.397.1836G}.
Our model is also similar to what is observed from supergiant HMXBs, although the power law is quite steep and the high-energy 
cutoff starts at a relatively low energy.

The archival \xmm\ spectrum from 2010 can be modelled by the same continuum, however requires a much higher absorption.
This indicates that \ngc\ was also in the ULX state at similar intrinsic X-ray luminosity, albeit highly absorbed in 2010. 
It is to be noted that the \cha\ spectra of \ngc\ from 2010 and 2014 exhibited a spectrum very similar to that observed with 
\xmm\ in 2010 \citep{2011ApJ...739L..51B,2016MNRAS.457.1636B}. The fact that we see a strong fluorescent Fe line in 2010, but 
not in 2016 is also consistent with the picture of very different absorption in these two epochs.
Statistical limitations of the earlier data inhibited a detailed spectral modelling in previous works. However, accounting for a 
highly absorbed continuum emission similar to that demonstrated in this paper can result in a much higher intrinsic luminosity 
than quoted in earlier publications. The high absorption during the first years after the supernova impostor event suggests that 
a large amount of material was expelled during the event which expanded and became now transparent to soft X-rays.

\section*{Acknowledgements}
This research has made use of data obtained with \xmm, an ESA science mission with instruments and contributions
directly funded by ESA Member States and NASA, and data obtained from the \nus\ Data Archive.
The \xmm\ project is supported by the Bundesministerium f{\"u}r Wirtschaft und Technologie/Deutsches Zentrum f{\"u}r Luft- und
Raumfahrt (BMWI/DLR, FKZ\,50OG1601) and the Max Planck Society.




\bibliographystyle{mnras}
 \bibliography{article}
 \label{lastpage}
 \end{document}